\documentclass[12pt,thmsa,a4paper,notitlepage]{article}
\usepackage{amssymb}

\usepackage{epsfig}

\begin{document}

\title{{\LARGE \textbf{QUANTUM REGISTER PHYSICS}}}
\author{\textbf{George Jaroszkiewicz} \\
School of Mathematical Sciences,\\
University of Nottingham, UK}
\maketitle

\begin{abstract}
Motivated by Feynman's 1983 paper on the simulation of physics by
computers, we present a general approach to the description of
quantum experiments which uses quantum bit registers to represent the
spatio-temporal changes occurring in apparatus-systems during the
course of such experiments. To illustrate our ideas, we discuss the
Stern-Gerlach experiment, Wollaston prisms, beam splitters,
Mach-Zender interferometers, von Neumann (PVM) tests, the more
general POVM formalism, and a variety of modern quantum experiments,
such as two-particle interferometry and the EPR scenario.
\end{abstract}

\section{Introduction}

It seems reasonable to state that at present, only about half of the laws of
physics are understood; whilst we know very well how to predict the outcome
probabilities of given quantum experiments, we have no idea as to why
we find ourselves doing those experiments in the first place. In
other words, we do not have a proper theory of the universe
considered as a fully autonomous, self-referential quantum dynamical
system.

In 1983, Feynman wrote a paper on the simulation of physics with computers
\cite{FEYNMAN-82}. Towards the end of the paper, he wrote:

\begin{quotation}
``...we have an illusion that we can do any experiment that we want. We all,
however, come from the same universe, have evolved with it, and don't really
have any ``real'' freedom. For we obey certain laws and have come from a
certain past.''
\end{quotation}

A number of points arise in connection with Feynman's paper, and with this
quotation in particular, which have motivated the work presented in the
present paper. First, Feynman's paper explores the notion that the laws of
physics, if not the universe as a whole, might be describable in terms of
computation. Because of quantum mechanics, however, Feynman recognized that
the computation involved could not be classical but has to involve what is
now known as quantum computing. Secondly, and this comes as a surprise
considering the pragmatic nature of Feynman's lifetime contributions to
quantum theory, Feynman seems to be advocating the study of \emph{%
endophysics }as opposed to \emph{exophysics. }Briefly, endophysics is
physics described from within, whilst exophysics is physics described from
the point of view of external observers looking into systems under
observation. The latter approach to physics has been very successful ever
since the time of Newton whilst the former remains a deep theoretical
challenge.

Given the success of exophysics, it seems at first sight unreasonable to
consider replacing it with an intractable alternative. However, there are no
signs that quantum mechanics has any natural boundaries. On the contrary,
there is increasing evidence for the applicability of quantum principles at
scales much greater than the atomic. The hypothetical line between the
classical and quantum worlds has been called the ``Heisenberg cut''. It does
not seem to exist. Sooner or later, we shall be forced to understand the
dynamics of observers in a more fundamental way, not just the dynamics of
the systems that those observers are looking at.

\

Given that we have been motivated to think of how observers and their
measuring apparatus evolve dynamically, we are faced with the challenge of
finding a dynamical description for them on a par with the quantum
description we have for systems under observation, such as the Schr\"{o}%
dinger equation. We are a long way from having anything like that, and it is
possible that such a goal might never be achieved. However, ruling out such
a possibility as a matter of principle seems a recipe for complacency, and
besides, could be a serious mistake.

In this paper, therefore, we attempt to bring into a quantum framework a
greater role for the physical apparatus involved in quantum experiments than
is usual. It will be evident, after reading our quantum register description
of experiments, what the limitations of our approach are. We present no
theory as to why experiments are done, but a start is made to bring into the
discussion some aspects involved in real quantum physics experiments which
generally have not been modelled in conventional approaches. Quantum
register physics has the potential to describe situations where not only
might physical apparatus change in time, but also those situations where
many independent or coupled quantum experiments are being conducted
simultaneously.

What is presented in this paper provides a consistent quantum
computational framework for the description of simple and complex
experiments, such as the Stern-Gerlach experiment, the Mach-Zender
interferometer, experiments conventionally requiring a POVM
description, and EPR -type experiments.

\

The plan of this paper is as follows. First we review the notions of quantum
bits and quantum registers. Then we start to discuss more and more complex
experiments from the point of view of quantum registers. Our aim here is to
show how real physics experiments can be successfully modelled in a novel
way which holds some promise of bringing physical apparatus into quantum
discussions. Finally, we shall discuss some of the conceptual ramifications
of quantum register physics.

\section{Quantum bits}

A \emph{classical bit }$\mathcal{B}$ is a system with two possible states: $%
|1)\equiv $`\emph{Yes}'$=\mathrm{`}True$'$=\mathrm{`}occupied$\textrm{' }and
$|0)\equiv `No$\textrm{'}$=`False$\textrm{'}$=`unoccupied$\textrm{'}. We may
represent these states by the two-dimensional real column vectors \textrm{\ }
\begin{equation}
|1)\equiv \left[
\begin{array}{c}
1 \\
0
\end{array}
\right] ,\;\;\;|0)\equiv \left[
\begin{array}{c}
0 \\
1
\end{array}
\right]
\end{equation}
and their duals $(1|$, $(0|$ by the row vectors
\begin{equation}
(1|\equiv \left[
\begin{array}{cc}
1 & 0
\end{array}
\right] ,\;\;\;(0|\equiv \left[
\begin{array}{cc}
0 & 1
\end{array}
\right] .
\end{equation}
Then we have the orthonormality condition $(i|j)=\delta _{ij}$ for \ any $%
i,j $ in the set $\left\{ 0,1\right\} .$

Given that $|\psi )$ is a classical bit state, but with no other
information, we may write\qquad
\begin{equation}
|\psi )=\alpha |1)+\beta |0),
\end{equation}
where $\alpha ,\,\beta $ are in the set $\left\{ 0,1\right\} $ and
\begin{equation}
(\psi |\psi )=|\alpha |^{2}+|\beta |^{2}=1.
\end{equation}
For a classical bit, there are only two possible sets of values for $(\alpha
,\beta )$, viz., $(1,0)$ or else $(0,1)$.

We turn a classical bit into a qubit (quantum bit) by regarding the
classical bit states $|0)$ and $|1)$ as the basis vectors \emph{\ }for a
2-dimensional Hilbert space $\mathcal{Q}$. These basis vectors will be
referred to as the \emph{computational basis}.

A general normalized qubit state is given by
\begin{equation}
|\psi )=\alpha |0)+\beta |1),
\end{equation}
where \ now $\alpha $, $\beta $ are allowed to be complex and satisfy the
normalization condition
\[
(\psi |\psi )=|\alpha |^{2}+|\beta |^{2}=1.
\]
Here $(0|$ and $(1|$ are the vectors dual to $|0|$ and $|1)$ respectively
and form the computational basis for the dual qubit Hilbert space $\mathcal{Q%
}^{\ast }$.

\section{Qubit operators}

For a given qubit $\mathcal{Q}$, we define the following operators:

\subparagraph{\textbf{i) the projection operators:}}

\begin{equation}
P^{0}\equiv |0)(0|,\;\;\;P^{1}\equiv |1)(1|
\end{equation}

\subparagraph{ii) \textbf{the transition operators}:}

\begin{equation}
A\equiv |0)(1|,\;\;\;A^{+}\equiv |1)(0|
\end{equation}

\subparagraph{\textbf{iii) the identity operator:}}

\begin{equation}
\sigma ^{0}\equiv P^{1}+P^{0}
\end{equation}

\subparagraph{\textbf{iv) the Pauli operators:}}

\begin{eqnarray}
&&\left.
\begin{array}{c}
\sigma ^{1}\equiv A+A^{+} \\
\;\sigma ^{2}\equiv iA-iA^{+}
\end{array}
\right\} \;\;\;\textrm{flip\ operators}\;\;\;\;\;\;\;\;  \nonumber \\
&&\;\;\,\,\sigma ^{3}\equiv P^{1}-P^{0}.
\end{eqnarray}

All of these operators apart from $\sigma ^{2}$ can be defined for classical
bits. Also, all of these operators can be multiplied together and form a
closed algebra, represented by Table 1. For example, the product $P^{1}A$ is
found by the intersection of the row labelled by $P^{1}$ and the column
labelled by $A$. From the table, we find $P^{1}A=0$, the zero operator.

This table turns out to be invaluable in quantum register physics. It should
be noted that it applies only to operators acting on the same qubit.\textbf{%
\ }

\begin{center}
\begin{tabular}[t]{|c|ccccccc|}
\hline
$^{\;}$ & $\mathbf{P}^{0}$ & $\mathbf{P}^{1}$ & $\mathbf{A}$ & $\mathbf{A}%
^{+}$ & $\mathbf{\sigma }^{1}$ & $\mathbf{\sigma }^{2}$ & $\mathbf{\sigma }%
^{3}$ \\ \hline
$\mathbf{P}^{0}$ & $P^{0}$ & $0$ & $A$ & $0$ & $A$ & $iA$ & $-P^{0}$ \\
$\mathbf{P}^{1}$ & $0$ & $P^{1}$ & $0$ & $A^{+}$ & $A^{+}$ & $-iA^{+}$ & $%
P^{1}$ \\
$\mathbf{A}$ & $0$ & $A$ & $0$ & $P^{0}$ & $P^{0}$ & $-iP^{0}$ & $A$ \\
$\mathbf{A}^{+}$ & $A^{+}$ & $0$ & $P^{1}$ & $0$ & $P^{1}$ & $iP^{1}$ & $%
-A^{+}$ \\
$\mathbf{\sigma }^{1}$ & $A^{+}$ & $A$ & $P^{1}$ & $P^{0}$ & $\sigma ^{0}$ &
$i\sigma ^{3}$ & $-i\sigma ^{2}$ \\
$\mathbf{\sigma }^{2}$ & $-iA^{+}$ & $iA$ & $-iP^{1}$ & $iP^{0}$ & $-i\sigma
^{3}$ & $\sigma ^{0}$ & $i\sigma ^{1}$ \\
$\mathbf{\sigma }^{3}$ & $-P^{0}$ & $P^{1}$ & $-A$ & $A^{+}$ & $i\sigma ^{2}$
& $-i\sigma ^{1}$ & $\sigma ^{0}$ \\ \cline{1-7}\cline{2-8}\cline{8-8}
\end{tabular}

\

Table $1.$ The computational basis operator algebra for a single qubit.
\end{center}

From Table $1$ we can read off the following fundamental property of the
transition operators: \qquad
\begin{equation}
AA=A^{+}A^{+}=0,  \label{AAA}
\end{equation}
which at first sight suggests that these operators are related to fermionic
or Grassmannian variables. However, qubits are neither spin-half fermions
nor Grassmannian variables \cite{WU+LIDAR-02}, but they can be used in the
manner of Jordan and Wigner \cite{JORDAN+WIGNER-28} to construct fermionic
quantum fields out of large collections (quantum registers) of qubits. The
nilpotency property (\ref{AAA}) of the transition operators gives quantum
register physics a very particular flavour, modelling the fact that a real
physics apparatus is either void (is not being used) or else can ``hold''
only one state at a time.

\section{Quantum registers}

A rank-$r$ quantum register $\mathcal{R}^{r}$ is the tensor product of $r$
distinct, labelled qubits:
\begin{equation}
\mathcal{R}^{r}\equiv \mathcal{Q}_{0}\otimes \mathcal{Q}_{1}\otimes \ldots
\otimes \mathcal{Q}_{r-1}.
\end{equation}
It is a complex Hilbert space of dimension $2^{r}$, containing \emph{%
separable }and \emph{entangled} states. This makes it ideal for discussing
quantum physics.

In the following, the left-right ordering of tensor products is not
significant, although labels are significant. For example, the rank-$2$
quantum registers $\mathcal{Q}_{0}\otimes \mathcal{Q}_{1}$ and $\mathcal{Q}%
_{1}\otimes \mathcal{Q}_{0}$ are equivalent.

A \emph{register computational basis} $B\left( \mathcal{R}^{r}\right) $ is
readily constructed by tensoring the computational bases for all of the
qubits in the register in the following manner:
\begin{equation}
B\left( \mathcal{R}^{r}\right) =\left\{ |i_{0})_{0}\otimes
|i_{1})_{1}\otimes \ldots \otimes |i_{r-1})_{r-1}:\left.
\begin{array}{c}
i_{j}\in \left\{ 0,1\right\} , \\
0\leqslant j\leqslant r-1.
\end{array}
\right. \right\} .\;\;\;\;\;\;\;\;\;\;
\end{equation}
In our work, we shall always denote qubit and register states by Dirac
bra-ket notation, modified by the replacement of angular brackets $\rangle
,\,\langle $ with round brackets $),($ respectively. Elements of the
register computational basis $B\left( \mathcal{R}^{r}\right) $ can be
represented in a number of equivalent ways, depending on context, as follows:

\begin{enumerate}
\item  We can drop the tensor product symbol, as the individual qubit
identifier labels suffice to carry the necessary information:
\begin{equation}
|i_{0})_{0}\otimes |i_{1})_{1}\otimes \ldots \otimes
|i_{r-1})_{r-1}=|i_{0})_{0}|i_{1})_{1}\ldots |i_{r-1})_{r-1};
\end{equation}

\item  We can write out register computational basis elements in terms of a
sequence of ones and zeros:
\begin{equation}
|i_{0})_{0}\otimes |i_{1})_{1}\otimes \ldots \otimes
|i_{r-1})_{r-1}=|i_{0}i_{1}i_{2}\ldots i_{r-1});
\end{equation}

\item  We can interpret such a sequence as a binary number and use that \
instead of the sequence:
\begin{equation}
|i_{0})_{0}\otimes |i_{1})_{1}\otimes \ldots \otimes
|i_{r-1})_{r-1}=|i_{0}2^{0}+i_{1}2^{1}+\ldots +i_{r-1}2^{r-1})
\end{equation}
Then the register computational basis can be written in the form
\begin{equation}
B\left( \mathcal{R}^{r}\right) =\left\{ |a):0\leqslant a\leqslant
2^{r}-1\right\} ,
\end{equation}
with orthonormality condition
\begin{equation}
(a|b)=\delta _{ab},\;\;\;\;\;0\leqslant a,b\leqslant 2^{r}-1.
\end{equation}
\end{enumerate}

For example, for a rank-$3$ quantum register, the element $|1)_{0}\otimes
|0)_{1}\otimes |1)_{2}$ can be written in the following ways:
\begin{equation}
|1)_{0}\otimes |0)_{1}\otimes
|1)_{2}=|1)_{0}|0)_{1}|1)_{2}=|101)=|1.2^{0}+0.2^{1}+1.2^{2})=|5).
\end{equation}
Occasionally, there will be possible ambiguity as to whether a number is
written in binary or in decimal. For instance, in a rank-$4$ register, we
have the state
\begin{equation}
|1101)=|1+2+8)=|11_{10})
\end{equation}
In such cases, we shall always give the decimal representation the subscript
$10$, to denote ``base ten'', and then we know that $|11_{10})=|``eleven")$
and not $|1.2^{0}+1.2^{1})=|3)$. Whenever there is no possible ambiguity, we
shall not need this subscript and so leave it out.

An arbitrary quantum register state $|\psi )$ can always be expressed
in terms of the register computational basis, i.e.,
\begin{equation}
|\psi )=\sum_{a=0}^{2^{r}-1}\psi _{a}|a),\;\;\;\;\;\psi _{a}\in \Bbb{C}.
\end{equation}
Then the inner product between any two elements $|\psi )$, $|\phi )$ of the
register is given by \
\begin{equation}
(\psi |\phi )=\sum_{a=0}^{2^{r}-1}\psi _{a}^{\ast }\phi _{a}.
\end{equation}

We are now ready to discuss real quantum physics experiments.

\section{The Stern-Gerlach experiment}

In 1922, Stern and Gerlach performed an experiment, passing electrons
through a strong, inhomogeneous magnetic field \cite
{STERN-GERLACH-22A,STERN-GERLACH-22B}. Their apparatus is represented in
Figure 1.

\

\begin{figure}[!h]
\centerline{\epsfxsize=3.0in \epsfbox{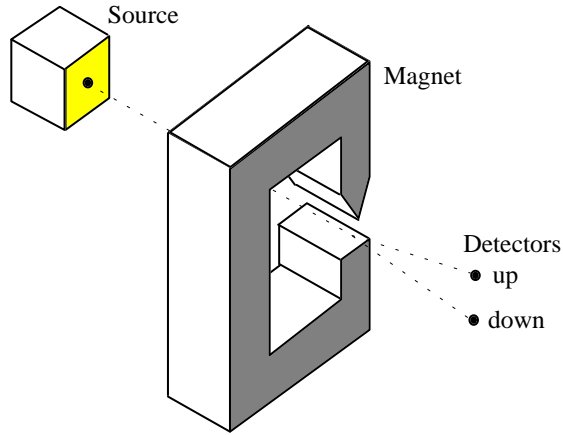}}
\caption{Idealized Stern-Gerlach experiment.}
\end{figure}

In the conventional Hilbert space description of this experiment, an
electron state evolves from an initial state $|\Psi _{in}\rangle $ to a
final state $|\Psi _{out}\rangle $ which can be represented as the linear
superposition of two possible outcome states, known as ``spin up'' and
``spin down'' respectively. Each of these states is associated with one of
the outcome spots shown in Figure 1. These outcomes are represented by the
orthonormalized kets $|$up$\rangle $ and $|$down$\rangle $ respectively:
\begin{equation}
|\Psi _{in}\rangle \rightarrow |\Psi _{out}\rangle =\alpha |\textrm{up}\rangle
+\beta |\textrm{down}\rangle, \;\;\;\;\; | \alpha |^2 + | \beta |^2
= 1.
\end{equation}
The statistical results of the experiment are in
agreement with the quantum theory Born probability rule
\begin{eqnarray} P\left( \textrm{up}|\Psi _{in}\right) &=&|\langle
\textrm{up}| \Psi _{out}\rangle |^{2}=|\alpha |^{2},  \nonumber \\
P\left( \textrm{down}|\Psi _{in}\right) &=&|\langle \textrm{down} |
\Psi _{out}\rangle |^{2}=|\beta |^{2}. \end{eqnarray}

Our quantum register description of this and other experiments rests on
several observations about what happens in such experiments.

First, in most real physics experiment, all parts of the apparatus exist
before, during and after the experiment. Normally, many individuals runs or
repetitions of such an experiment are performed, generally spaced over a
significant interval of time and, during this time, the apparatus maintains
a temporally enduring identity. We shall model this aspect of the physics in
our quantum register description.

Second, for the Stern-Gerlach experiment illustrated in Figure $1$, the two
outcome possibilities for each emerging electron are detected at \emph{%
different} spatial locations (of course, in any single run involving a
single electron passing through the apparatus, the electron is detected at
only one of these two places at the end of that run). This spatial
separation is crucial to the success of the experiment, for without it,
Stern and Gerlach would never have been able to observe anything unusual.
This aspect of detection is also modelled in our quantum register physics.

Third, it is possible to have more than one Stern-Gerlach experiment being
performed simultaneously in different parts of the world. The quantum
register description permits us to describe such a scenario using a single
large quantum register.

\subsection{The quantum register description:}

The essence of the quantum register description of the Stern-Gerlach and
other experiments is to assign a qubit to each place where physicists could
in principle detect new information. This is equivalent to using the \emph{%
space} concept in a counterfactual way. For the Stern-Gerlach experiment,
this means assigning at least three qubits as follows:

\

1) we assign a qubit $\mathcal{Q}_{0}$ to the source of the electrons;

2) we assign a qubit $\mathcal{Q}_{1}$ to the \emph{up} state detector;

3) we assign a qubit $\mathcal{Q}_{2}$ to the \emph{down }state detector.

\

This assignment is represented in Figure 2:

\

\begin{figure}[!h]
\centerline{\epsfxsize=3.0in \epsfbox{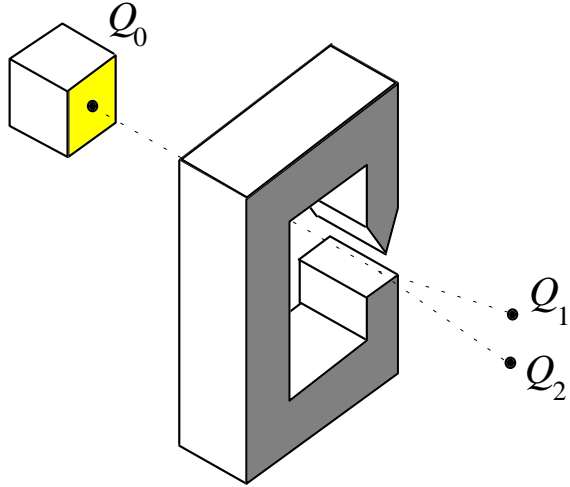}}
\caption{Stern-Gerlach experiment qubit assignment.}
\end{figure}

Points to note are

\begin{enumerate}
\item  Unlike the conventional description of the Stern-Gerlach experiment,
states $|$up$\rangle $ and $|$down$\rangle $ are \emph{not} regarded as
orthogonal qubit states in the same qubit space;

\item  We include the source in the description;

\item  In principle we could imagine the physical space between source and
detectors as filled with qubits, but these would be redundant here. We need
only that number of qubits sufficient to model the essential physics of a
given experiment. Later on in this paper we shall discuss quantum register
cosmology and quantum register field theory, where there is a reason to
consider space in terms of a very large (possibly infinite) rank quantum
register.
\end{enumerate}

Having set up a rank-three quantum register \ $\mathcal{R}^{3}\equiv
\mathcal{Q}_{0}\otimes \mathcal{Q}_{1}\otimes \mathcal{Q}_{2}$ to model the
Stern-Gerlach experiment, we now discuss a typical run of this experiment.
This involves a time-dependent description of the state $|\Psi )$ of the
combined apparatus-system.

First, imagine the situation \emph{after }all apparatus has been constructed
but before the actual experiment has started. During such a time, the
equipment is lying idle, i.e., unused. It exists, but no electron is being
prepared and no detector is registering any result. Such a state of the
apparatus-laboratory system will be called the \emph{void} state (we shall
not use the term $vacuum$ in this context, as this will be reserved for
other specific situations). We shall represent the void state by the quantum
register vector
\begin{equation}
\fbox{$|\Psi _{0})=|0)_{0}|0)_{1}|0)_{2}=|000)=|0)$}.
\end{equation}

Suppose now that the experiment has started. At some initial time $t_{in}$,
the experimentalists will be confident that the source has prepared an
initial state, but nothing has yet been registered by either detector. We
represent the lab-state (our terminology for the state of the apparatus and
system) by the quantum register state
\begin{equation}
\fbox{$|\Psi _{in})=|1)_{0}|0)_{1}|0)_{2}=|100)=|1)=\Bbb{A}_{0}^{+}|0)$},
\end{equation}
where in this particular case
\begin{equation}
\Bbb{A}_{0}^{+}\equiv A_{0}^{+}\otimes \sigma _{1}^{0}\otimes \sigma
_{2}^{0}.
\end{equation}
Note that we can be sure that there must be such an interval of time,
because the detectors are spatially separated from the source, and therefore
could only trigger a non-zero time \emph{after }state preparation, according
to the principles of special relativity. It does not matter that according
to some quantum theorists, quantum states change instantaneously.
This is irrelevant in quantum register physics. All signals registered
in our qubits have to be consistent with relativity.

Finally, at a time $t_{out}>t_{in}$, we may write down the lab-state
immediately prior to detection:
\begin{equation}
\fbox{$|\Psi _{out})=\alpha |010)+\beta |001)=\alpha |2)+\beta |4)=\left(
\alpha \Bbb{A}_{1}^{+}+\beta \Bbb{A}_{2}^{+}\right) |0)$},
\end{equation}
where
\begin{equation}
\Bbb{A}_{1}^{+}\equiv \sigma _{0}^{0}\otimes A_{1}^{+}\otimes \sigma
_{2}^{0},\;\;\;\Bbb{A}_{2}^{+}\equiv \sigma _{0}^{0}\otimes \sigma
_{1}^{0}\otimes A_{2}^{+}.
\end{equation}
Once we have determined $|\Psi _{out}),$ the Born probability rule adapted
to the quantum register can be applied to give the outcome probabilities
\begin{eqnarray}
P\left( \textrm{up}|\Psi _{in}\right) &\equiv &|(2|\Psi _{out})|^{2}=|\alpha
|^{2}  \nonumber \\
P\left( \textrm{down}|\Psi _{in}\right) &\equiv &|(4|\Psi _{out})|^{2}=|\beta
|^{2},  \nonumber \\
P\left( \textrm{any other state} | \Psi _{in}\right) &\equiv &|\left( a|\Psi
_{out}\right) |^{2}=0,\;\;\;a=0,1,3,5,6,7,
\end{eqnarray}
consistent with known physics. Of course, during any single run involving a
single electron, only one detector gets triggered, so these probabilities
have to be related to the frequencies of outcome built up over many runs of
the basic experiment.

Normally, after each run is over and before the next one starts, the
lab-state reverts to the void state $|0)$. The specific mechanism for this
is currently beyond known physics, as is the transition from the void state
to the initial state at the start of a run.

Our quantum register approach permits a very interesting and physically
meaningful situation to be contemplated. We could imagine starting the $%
\left( n+1\right) ^{th}$ run \emph{before }the\emph{\ }$n^{th}$ is complete.
Such a situation would be modelled, for example, by the sequence
\begin{equation}
|\Psi _{n})\equiv \alpha |010)+\beta |001)\rightarrow |\Psi _{n+1})\equiv
\alpha |110)+\beta |101).
\end{equation}
We shall discuss this and other exotic possibilities in $\S 12$.

\

More formally, we may consider changes in the lab-state to be described by
unitary evolution over $\mathcal{R}^{3}$, viz.,
\[
|\Psi _{in})\rightarrow |\Psi _{out})\equiv \Bbb{U}\left(
t_{out},t_{in}\right) |\Psi _{in}),
\]
where $\Bbb{U}\left( t_{out},t_{in}\right) $ is unitary so as to preserve
total probability. Exactly what this operator is or should be will not
always be clear, because physics experiments will not in general deal with
absolutely every possible state in a quantum register. The basic
Stern-Gerlach experiment, for example, requires us only to consider four of
the eight computational basis elements, viz, $|0),|1),|2)$ and $|4)$. For
such basic systems, there is a degree of overkill in the quantum register
description. This should not be regarded as a flaw; a similar situation
occurs in most classical and quantum theories.

We can be sure of one or two things, however. First, if the laboratory is in
a void state, then we do not expect that to change, unless we initiate a new
run (which will not conserve probability anyway). Therefore, we may assume
\begin{equation}
\Bbb{U}\left( t_{out},t_{in}\right) |0)=|0).
\end{equation}
Then we can represent the dynamics in terms of how the transition
operators change, viz.,
\begin{equation}
\Bbb{A}_{0}^{+}\rightarrow \Bbb{U}\left( t_{out},t_{in}\right) \Bbb{A}%
_{0}^{+}\Bbb{U}^{+}\left( t_{out},t_{in}\right) =\alpha \Bbb{A}%
_{1}^{+}+\beta \Bbb{A}_{2}^{+}.
\end{equation}
More generally, we shall ``modularise'' our spatio-temporal description,
meaning that individual transition operators will change at various times in
their own ways. Typically, we shall leave out specific reference to
the $\Bbb{U}$ operators,  writing for example
\begin{equation}
\Bbb{A}_{0}^{+}\rightarrow \alpha \Bbb{A}_{1}^{+}+\beta \Bbb{A}_{2}^{+}
\label{CCC}
\end{equation}
to describe a particular change in the operator $\Bbb{A}_{0}^{+}$ at a
particular place and time during a given run.

Two important points need to be noted. First, our quantum register
description is not designed to give us the dynamical details of such
transitions. For that we need to invoke standard quantum mechanics. The
quantum register description is designed to show more consistently how
sequences of such dynamical changes get distributed around in time and
space, making overall calculations of complex processes easier to calculate.

Second, because real physics experiments are irreversible, we need to be
cautious about what operators such as $\Bbb{U}\left( t_{out},t_{in}\right) $
really mean. They will have the semi-group property
\begin{equation}
\Bbb{U}\left( t_{2},t_{1}\right) \Bbb{U(}t_{1},t_{0}\Bbb{)=U(}t_{2},t_{0}%
\Bbb{)},\;\;\;t_{2}\geqslant t_{1}\geqslant t_{0}
\end{equation}
and satisfy
\begin{equation}
\Bbb{U}\left( t_{1},t_{0}\right) \Bbb{U}^{+}\left( t_{1},t_{0}\right) =\Bbb{I%
}_{\mathcal{R}},
\end{equation}
where $\Bbb{I}_{\mathcal{R}}$ is the register identity operator, but we may
have no clear physical interpretation of what the operator
$\Bbb{U}\left( t_{0},t_{1}\right) $ means, for $t_{0}<t_{1}$.

\section{The Wollaston prism}

The two polarization degrees of freedom of photons make them analogous to
electrons in certain situations. For instance, the passage of an
monochromatic electromagnetic wave through a Wollaston prism splits the wave
into two spatially distinct waves, identified with two distinct mutually
orthogonal transverse polarization components, shown in Figure 3.

\begin{figure}[!t]
\centerline{\epsfxsize=2.0in \epsfbox{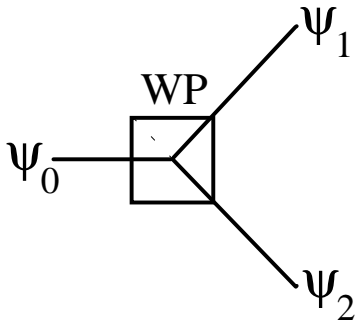}}
\caption{Passage of monochromatic light through a Wollaston prism.}
\end{figure}

If $|\psi _{0}\rangle $ is a monochromatic photon state, we may write
\begin{equation}
|\psi _{0}\rangle =\psi _{1}|x\rangle +\psi _{2}|y\rangle ,
\end{equation}
where $|x\rangle $ and $|y\rangle $ represent the two mutually orthogonal
transverse polarization vectors involved.

The quantum register description of a Wollaston prism, shown in Figure 4,
turns out to be identical to that for the Stern-Gerlach experiment. In
operator terms, we find
\begin{equation}
\Bbb{A}_{0}^{+}\rightarrow \psi _{1}\Bbb{A}_{1}^{+}+\psi _{2}\Bbb{A}_{2}^{+},
\end{equation}
which is formally identical to (\ref{CCC}).

\

\

\begin{figure}[!h]
\centerline{\epsfxsize=1.5in \epsfbox{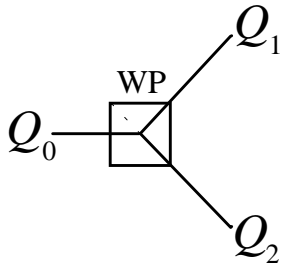}}
\caption{Quantum register description of a Wollaston prism.}
\end{figure}

\section{von Neumann tests}

The Stern-Gerlach and Wollaston prism experiments are the most elementary
and useful examples of the sort of quantum experiments discussed by von
Neumann \cite{VON-NEUMANN:55}, where an ensemble of identically prepared
initial states is passed through some test apparatus $A$ and a range of
possible outcomes detected. The description of an idealized version of such
an experiment leads to the so-called \emph{projection valued measure} (PVM)
description of quantum experiments. This is known to have its limitations,
but remains an important concept.

The general PVM test is shown in Figure $5$.
For each run of an ensemble of runs, the initial state $|\Psi_{in}\rangle $,
which will be assumed to be pure, is prepared by some apparatus $\Sigma _{0}$
at time $t_{in}$. Subsequently, the prepared state is passed through test
apparatus $A$, and one out of $d$ possible outcomes detected at time $%
t_{out} $. In von Neumann's approach, $|\Psi _{in}\rangle $ is assumed to be
a normalized element of some $d-$dimensional Hilbert space $\mathcal{H}$.
The test $A$ is represented by some non-degenerate Hermitian operator $\hat{A%
}$ acting over $\mathcal{H}$. Because of non-degeneracy, the $eigenstates$ $%
|a_{1}\rangle $, $|a_{2}\rangle $,$\ldots ,|a_{d}\rangle $ can be normalized
and form an orthonormal basis for $\mathcal{H}$, known as the \emph{%
preferred basis.}

\

\begin{figure}[!h]
\centerline{\epsfxsize=2.5in \epsfbox{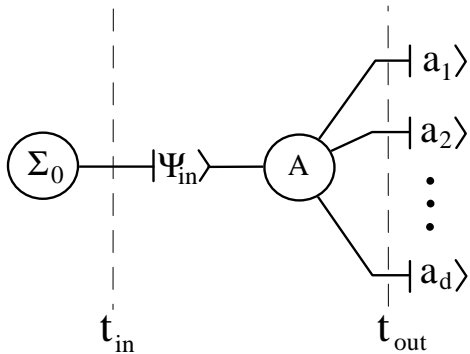}}
\caption{A general PVM, or von Neumann, test.}
\end{figure}

Because of completeness, we may write
\begin{equation}
|\Psi _{in}\rangle \rightarrow |\Psi _{out}\rangle =\hat{U}\left(
t_{out},t_{in}\right) |\Psi _{in}\rangle =\sum_{i=1}^{d}\Psi
^{i}|a_{i}\rangle ,
\end{equation}
where
\begin{equation}
\Psi ^{i}=\langle a_{i}|\Psi _{out}\rangle =\langle a_{i}|\hat{U}\left(
t_{out},t_{in}\right) |\Psi _{in}\rangle .  \label{DDD}
\end{equation}
The Born probability interpretation then predicts the conditional outcome
probabilities to be given by
\begin{equation}
P\left( a_{i}|\Psi _{in}\right) =|\langle a_{i}|\Psi _{out}\rangle
|^{2}=|\Psi ^{i}|^{2}.
\end{equation}

The quantum register description of a PVM scenario follows the pattern
outlined for the Stern-Gerlach and Wollaston prism experiments. We associate
one qubit with every part of the apparatus wherever a state could be
detected and new information acquired. This means one qubit for the
preparation apparatus and one for each of the $d$ possible outcomes, shown
in Figure $6$. Therefore, we need a rank-$\left(1+d\right) $ quantum
register for such a test.

\

\begin{figure}[!h]
\centerline{\epsfxsize=2.0in \epsfbox{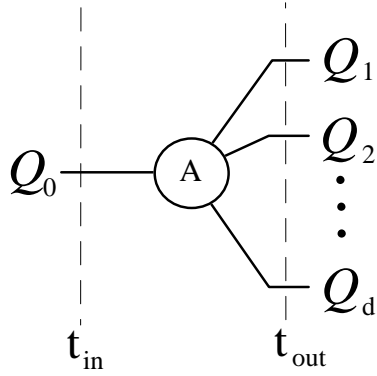}}
\caption{Qubit assignment for a general PVM, or von Neumann, test.}
\end{figure}

The quantum dynamics is given by the rule
\begin{equation}
\Bbb{A}_{0}^{+}\rightarrow \sum_{i=1}^{d}\Psi ^{i}\Bbb{A}_{i}^{+},
\end{equation}
where the $\Psi ^{i}$ are given by the conventional quantum
calculation (\ref {DDD}), so we find
\begin{equation}
|\Psi _{in}\rangle \rightarrow |\Psi _{out})=\sum_{i=1}^{d}\Psi ^{i}\Bbb{A}%
_{i}^{+}|0)=\sum_{i=1}^{d}\Psi ^{i}|2^{i}).
\end{equation}
The conditional probabilities for the $d$ possible outcomes of the
experiment are then given by the quantum register Born rule
\begin{equation}
P\left( a_{i}|\Psi _{in}\right) \equiv |\left( 2^{i}|\Psi _{out}\right)
|^{2}=|\Psi ^{i}|^{2},
\end{equation}
as before.

If all experiments were of this form, there would be little advantage in the
quantum register description. This comes into its own when more than one von
Neumann test are coupled together, a situation which occurs frequently in
quantum optics experiments.

An important observation about the formalism developed thus far is that all
physical states so far considered are linear combinations of only certain
elements of the computational basis, viz, those of the form
\begin{equation}
|2^{k})\equiv \Bbb{A}_{k}^{+}|0).
\end{equation}
States of the form $|2^{k})$ and linear combinations of such states, will be
called \emph{rank-one states. }We define\emph{\ }rank-$p$ states to be
linear combinations of elements of the computational basis given by
\begin{equation}
\Bbb{A}_{i_{1}}^{+}\Bbb{A}_{i_{2}}^{+}\ldots
\Bbb{A}_{i_{p}}^{+}|0)=|2^{i_{1}}+2^{i_{2}}+\ldots +2^{i_{p}})
\end{equation}
where the $0\leqslant i_{j}<i_{j+1}\leqslant r-1$. For example,
\begin{equation}
\Bbb{A}_{2}^{+}\Bbb{A}_{3}^{+}\Bbb{A}_{5}^{+}|0)=|001101000...0)=|2^{2}+2^{3}+2^{5})=|44)
\end{equation} is a rank-$3$ state.

\section{More general experiments}

Eventually, it become apparent that the PVM formulation of quantum physics
is too limited and it became superseded by the more general POVM (Positive
Operator Valued Measure) approach. In this new approach, quantum experiments
can have more or less outcomes than the dimension of the Hilbert space
involved. For example, suppose we have an experiment with $k$ possible
outcomes, with $k$ not necessarily equal to $d$, the dimension of the
Hilbert space $\mathcal{H}$ used to model the states of the system. For each
outcome $|\phi ^{i}\rangle $, $i=1,2,\ldots ,k$, there is an associated
positive operator $\hat{E}_{i}$, such that
\begin{equation}
\sum_{i=1}^{k}\hat{E}_{i}=\hat{I}_{\mathcal{H}},  \label{BBB}
\end{equation}
where $\hat{I}_{\mathcal{H}}$ is the identity operator over the Hilbert
space. Given a normalized initial state $|\Psi \rangle \in \mathcal{H}$,
then the conditional probability $P\left( \phi ^{i}|\Psi \right) $ of
outcome $|\phi ^{i}\rangle $ is given by
\begin{equation}
P\left( \phi ^{i}|\Psi \right) =\langle \Psi |\hat{E}^{i}|\Psi \rangle ,
\end{equation}
with condition (\ref{BBB}) ensuring probabilities sum to unity.

The above discussion involves pure states. In fact, the POVM approach is
more general than this and can be extended to cover mixed states, which
requires a density matrix approach involving the taking of traces. We shall
not be interested in this paper in such situations. The generalization of
our quantum register description to cover such cases is not anticipated to
be particularly difficult and is left for future consideration.

We shall discuss now some situations where the POVM approach would normally
be invoked and give an alternative quantum register description. We recall
that an arbitrary POVM with a finite number of elements can always be
converted into a von Neumann (maximal) test by the introduction of an
auxiliary, independently prepared quantum system known as an ancilla, a
result which utilizes Neumark's theorem \cite{PERES:93}. Essentially, the
original Hilbert space $\mathcal{H}$ is extended into one of higher
dimension, $\mathcal{H}^{\prime }$, and von Neumann's PVM formulation can be
applied to $\mathcal{H}^{\prime }$ directly.

The disadvantage of this approach is that it masks the spatio-temporal
structure of the measurements involved and suggests that the simple $in-out$
structure of a single von Neumann test is all that is going on. In reality,
complex experiments involve sequences of processes rather like what happens
in a computer, which is why quantum computation is one possible way to
approach physics \cite{FEYNMAN-82}.

Suppose for example that instead of irreversibly registering all information
about the outcomes of a von Neumann test, we feed one or more of its
possible outcome channels into some new test. Quantum interference
experiments, such as double-slit and Mach-Zender interferometer experiments,
are of this form. As an example, consider the double Stern-Gerlach
experiment shown in Figure $7.$ An electron is prepared by apparatus $\Sigma
_{0}$ as in the original Stern-Gerlach experiment and passed through a
Stern-Gerlach apparatus $SG_{1}$ which has quantization axis \ along vector $%
\mathbf{k.}$ Any spin down outcome $|-\mathbf{k}\rangle $ is recorded,
whereas any spin up outcome $|+\mathbf{k}\rangle $ \ is not registered but
channelled into a second Stern-Gerlach apparatus $SG_{2},$ which has
quantization axis $\mathbf{a}$, with each of its possible outcomes being
detectable.

\

\

\begin{figure}[!h]
\centerline{\epsfxsize=4.0in \epsfbox{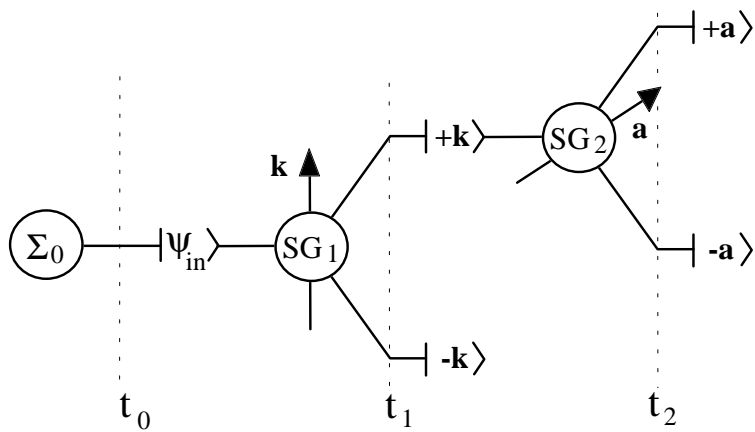}}
\caption{A double Stern-Gerlach experiment.}
\end{figure}

For any single run of the combined experiment, there are now three possible
mutually exclusive outcomes, viz., $|+\mathbf{a\rangle ,|-a\rangle }$ and $|-%
\mathbf{k\rangle }$, not two. For such an experiment, a PVM description
involving a Hilbert space of dimension two is not adequate. Conventionally,
either a POVM description with three positive operators over a
two-dimensional Hilbert space is needed, or an ancilla has to be introduced
if a PVM approach is desired, but this then requires an extension of the
original Hilbert space.

The alternative we propose is a quantum register description, which
describes such experiments quite readily. For the particular experiment
shown in Figure $7$, we require a rank-$5$ quantum register, as shown in
Figure $8$.

In this experiment, the outcome channel $|+\mathbf{k\rangle }$ of test $%
SG_{1}$ serves as an \emph{ideal measurement} \cite{PERES:93} or preparation
for $SG_{2}$; it is not absorbed by the detector but is used as an initial
state for the subsequent test $SG_{2}$. This seems to be the only physically
meaningful interpretation of the concept of ``state reduction''. State
reduction without subsequent testing is physically meaningless. Therefore,
rather than represent a final stage in a quantum process, state reduction
should always be considered as a beginning.

\

\

\begin{figure}[!h]
\centerline{\epsfxsize=4.0in \epsfbox{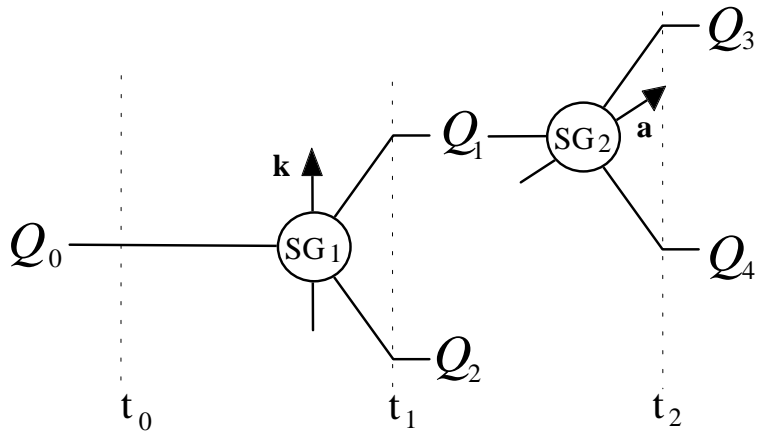}}
\caption{A qubit assignment for the double Stern-Gerlach experiment shown in
Figure 7.}
\end{figure}

\section{Beam splitters}

In optics, a beam splitter acts as a semi-transparent mirror, whereby part
of an incident electromagnetic wave is reflected and part transmitted, as
shown in Figure $9a$:

In quantum optics, such a device is usually regarded as having \emph{two}
input ports and two output ports, as in Figure $9b$, and is used in
experiments involving quantum interference, such as the Mach-Zender
experiment, discussed below.

We recover the single input channel picture when one of the two input
channels acts as a \emph{vacuum} or void port. More generally, for a
lossless beam splitter, the input and output waves are consistent with
unitary evolution, and can be written in the form \cite{ZEILINGER-81}:
\begin{equation}
\left[
\begin{array}{c}
\psi _{3} \\
\psi _{4}
\end{array}
\right] =e^{i\eta }\left[
\begin{array}{cc}
a & b \\
-b^{\ast } & a^{\ast }
\end{array}
\right] \left[
\begin{array}{c}
\psi _{1} \\
\psi _{2}
\end{array}
\right]
\end{equation}
where
\begin{equation}
|a|^{2}+|b|^{2}=1
\end{equation}
and $\eta $ is real. Then
\begin{equation}
|\psi _{3}|^{2}+|\psi _{4}|^{2}=|\psi _{1}|^{2}+|\psi _{2}|^{2}.
\end{equation}
\

\

\begin{figure}[!t]
\centerline{\epsfxsize=4.0in \epsfbox{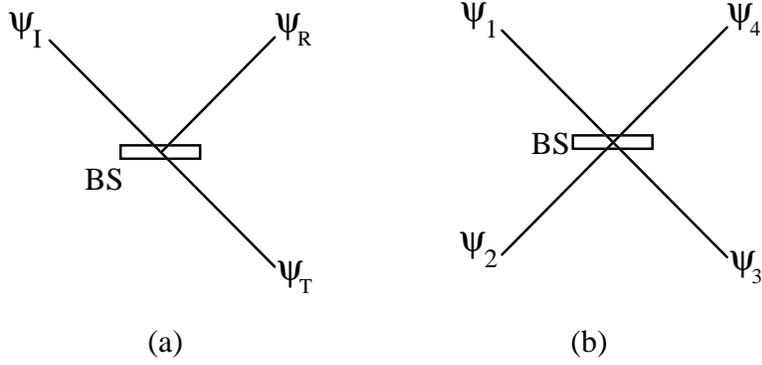}}
\caption{(a) The action of a beam splitter on a single incident
electromagnetic wave, (b) a beam splitter with two in-ports is used in
quantum interference experiments.}
\end{figure}

A quantum register description of a beam splitter requires four
qubits, as in Figure $10:$

\

\

\begin{figure}[!h]
\centerline{\epsfxsize=2.0in \epsfbox{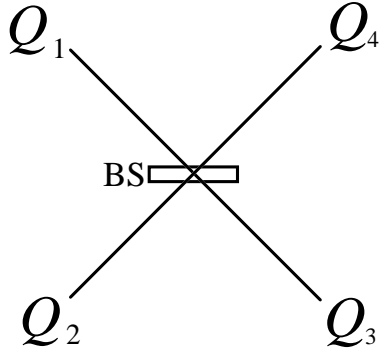}}
\caption{Qubit description of a beam splitter.}
\end{figure}

The correct temporal evolution is given by
\begin{eqnarray}
\Bbb{A}_{1}^{+} &\rightarrow &e^{i\eta }\left\{ a\Bbb{A}_{3}^{+}-b^{\ast }%
\Bbb{A}_{4}^{+}\right\}  \nonumber \\
\Bbb{A}_{2}^{+} &\rightarrow &e^{i\eta }\left\{ b\Bbb{A}_{3}^{+}+a^{\ast }%
\Bbb{A}_{4}^{+}\right\} .
\end{eqnarray}

This analysis applies to rank-1 states as well as rank-2 states. Figure $10$
on its own suggests that a rank-2 initial state is involved, such as
\begin{equation}
\Bbb{A}_{1}^{+}\Bbb{A}_{2}^{+}|0),
\end{equation}
but the analysis applies equally to rank -1 initial states such as
\begin{equation}
\left\{ \Bbb{A}_{1}^{+} + \Bbb{A}_{2}^{+}\right\}|0),
\end{equation}
and it is the latter which are involved in quantum interference usually.

\section{The Mach-Zender interferometer}

We are now in a position to consider more complex experiments via our
quantum register formalism. First, we shall discuss the Mach-Zender
interferometer, shown in Figure $11.$ A monochromatic beam of light $\Psi
_{0}$ is incident on a beam splitter $BS_{1}$, with output channels $\Psi
_{1}$, $\Psi _{2}$. The latter channel is passed through a device giving a
phase-shift $\phi $. Beams $\Psi _{2}$ and $\Psi _{3}$ are then deflected by
mirrors $M_{1}$, $M_{2}$ onto a second beam-splitter $BS_{2}$, identical to $%
BS_{1}$, and finally, its output channels lead on to photon detectors $D_{1}$%
, $D_{2}$.

\

\

\begin{figure}[!h]
\centerline{\epsfxsize=4.0in \epsfbox{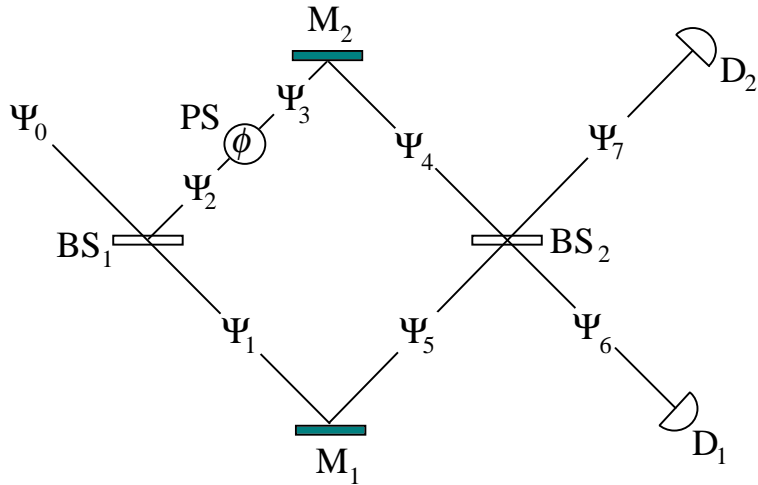}}
\caption{A Mach-Zender interferometer.}
\end{figure}

Taking the individual modules of the apparatus in turn, a conventional
wave-function description goes as follows:

\subparagraph{i) Beam splitter $BS _{1}$:}

\begin{equation}
\left[
\begin{array}{c}
\psi _{0} \\
0
\end{array}
\right] \rightarrow \left[
\begin{array}{c}
\psi _{1} \\
\psi _{2}
\end{array}
\right] =e^{i\eta }\left[
\begin{array}{cc}
a & b \\
-b^{\ast } & a^{\ast }
\end{array}
\right] \left[
\begin{array}{c}
\psi _{0} \\
0
\end{array}
\right]
\end{equation}

\subparagraph{ii) Phase shift $\protect\phi $:}

\begin{equation}
\left[
\begin{array}{c}
\psi _{1} \\
\psi _{2}
\end{array}
\right] \rightarrow \left[
\begin{array}{c}
\psi _{1} \\
\psi _{3}
\end{array}
\right] =\left[
\begin{array}{cc}
1 & 0 \\
0 & e^{i\phi }
\end{array}
\right] \left[
\begin{array}{c}
\psi _{1} \\
\psi _{2}
\end{array}
\right]
\end{equation}

\subparagraph{iii) Mirrors $M _{1}$, $M_{2}$:}

\begin{equation}
\left[
\begin{array}{c}
\psi _{1} \\
\psi _{3}
\end{array}
\right] \rightarrow \left[
\begin{array}{c}
\psi _{4} \\
\psi _{5}
\end{array}
\right] =\left[
\begin{array}{cc}
0 & e^{i\mu } \\
e^{i\mu } & 0
\end{array}
\right] \left[
\begin{array}{c}
\psi _{1} \\
\psi _{3}
\end{array}
\right] ,
\end{equation}
where we assume some phase shift $\mu $ due to reflection;

\subparagraph{iv) Beam splitter $BS_{2}$:}

\begin{equation}
\left[
\begin{array}{c}
\psi _{4} \\
\psi _{5}
\end{array}
\right] \rightarrow \left[
\begin{array}{c}
\psi _{6} \\
\psi _{7}
\end{array}
\right] =e^{i\eta }\left[
\begin{array}{cc}
a & b \\
-b^{\ast } & a^{\ast }
\end{array}
\right] \left[
\begin{array}{c}
\psi _{4} \\
\psi _{5}
\end{array}
\right]
\end{equation}

The result is that the waves incident on detectors $D_{1}$ and $D_{2}$ are
given by

\begin{eqnarray}
\psi _{6} &=&e^{i\left( 2\eta +\mu \right) }\left[ ab-e^{i\phi }ab^{\ast }%
\right] \psi _{0}  \nonumber \\
\psi _{7} &=&e^{i\left( 2\eta +\mu \right) }\left[ |a|^{2}+e^{i\phi }\left(
b^{\ast }\right) ^{2}\right] \psi _{0}
\end{eqnarray}

The quantum register description follows the prescription used before, which
is to place a qubit at every place where a quantum measurement/observation
\emph{could} in principle extract new information, as shown in Figure 12:

\

\

\begin{figure}[!h]
\centerline{\epsfxsize=4.0in \epsfbox{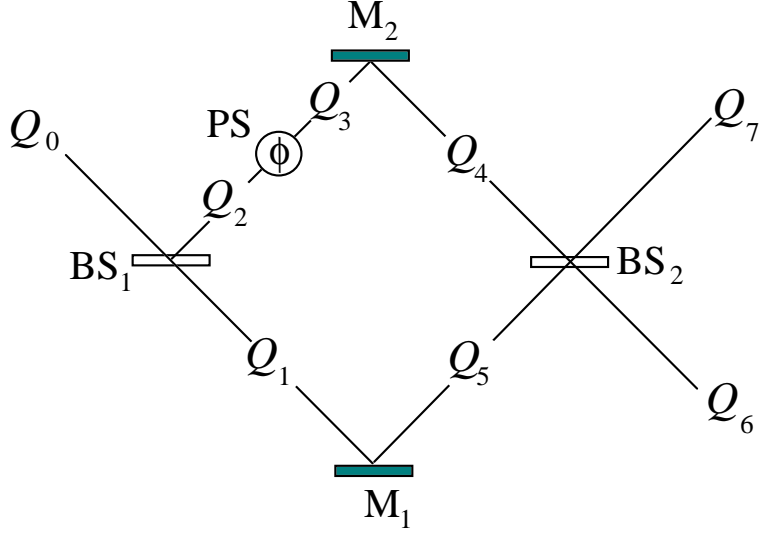}}
\caption{Qubit assignment for a Mach-Zender interferometer.}
\end{figure}

This suggests we need at least a rank-$8$ quantum register. However, if we
ignored the effect of the mirrors, which simply give an unobservable change
of phase in the overall amplitude, we could make do with two qubits less.
The quantum register calculation goes as follows:

\subparagraph{i) Beam splitter $BS_{1}$:}

\begin{equation}
\Bbb{A}_{0}^{+}\rightarrow e^{i\eta }a\Bbb{A}_{1}^{+}-e^{i\eta }b^{\ast }%
\Bbb{A}_{2}^{+},
\end{equation}

\subparagraph{ii) Phase shift $\protect\phi$:}

\begin{equation}
\Bbb{A}_{2}^{+}\rightarrow e^{i\phi }\Bbb{A}_{3}^{+},
\end{equation}

\subparagraph{iii) Mirrors $M_{1}$, $M_{2}$:}

\begin{equation}
\Bbb{A}_{1}^{+}\rightarrow e^{i\mu }\Bbb{A}_{5}^{+},\;\;\;\Bbb{A}%
_{3}^{+}\rightarrow e^{i\mu }\Bbb{A}_{4}^{+},
\end{equation}

\subparagraph{iv) Beam splitter $BS_{2}$:}

\begin{eqnarray}
\Bbb{A}_{4}^{+} &\rightarrow &e^{i\eta }a\Bbb{A}_{6}^{+}-e^{i\eta }b^{\ast }%
\Bbb{A}_{7}^{+}  \nonumber \\
\Bbb{A}_{5}^{+} &\rightarrow &e^{i\eta }b\Bbb{A}_{6}^{+}+e^{i\eta }a^{\ast }%
\Bbb{A}_{7}^{+}
\end{eqnarray}
Hence the register dynamics gives
\begin{eqnarray}
\Bbb{A}_{0}^{+} &\rightarrow &e^{i\left( 2\eta +\mu \right) }\left[
ab-e^{i\phi }ab^{\ast }\right] \Bbb{A}_{6}^{+}  \nonumber \\
&&+e^{i\left( 2\eta +\mu \right) }\left[ |a|^{2}+e^{i\phi }\left( b^{\ast
2}\right) \right] \Bbb{A}_{7}^{+},
\end{eqnarray}
i.e.,
\begin{eqnarray}
|\psi _{in})\equiv \Bbb{A}_{0}^{+}|0)\rightarrow |\psi _{out}) &=&\left\{
\begin{array}{c}
e^{i\left( 2\eta +\mu \right) }\left[ ab-e^{i\phi }ab^{\ast }\right] \Bbb{A}%
_{6}^{+} \\
+e^{i\left( 2\eta +\mu \right) }\left[ |a|^{2}+e^{i\phi }\left( b^{\ast
2}\right) \right] \Bbb{A}_{7}^{+}
\end{array}
\right\} |0)  \nonumber \\
&=&e^{i\left( 2\eta +\mu \right) }\left[ ab-e^{i\phi }ab^{\ast }\right]
|2^{6}) \\
&&+e^{i\left( 2\eta +\mu \right) }\left[ |a|^{2}+e^{i\phi }\left( b^{\ast
2}\right) \right] |2^{7}).  \nonumber
\end{eqnarray}
Note that the final state is still a rank-1 state. The amplitudes at the
detectors are given by
\begin{eqnarray}
\textrm{at }D_{1} &:&\;\;\;\;\;(2^{6}|\psi _{out})=e^{i\left( 2\eta +\mu
\right) }\left[ ab-e^{i\phi }ab^{\ast }\right]  \nonumber \\
\textrm{at }D_{2} &:&\;\;\;\;\;(2^{7}|\psi _{out})=e^{i\left( 2\eta +\mu
\right) }\left[ |a|^{2}+e^{i\phi }\left( b^{\ast 2}\right) \right]
\end{eqnarray}
in precise agreement with the conventional calculation shown earlier.

\section{Quantum interference POVM example}

We now consider a more complex experiment discussed recently, which
requires a POVM description \cite{BRANDT-99,BRANDT-02}. In
this experiment, a photon beam first passes through a Wollaston prism
and its output channels pass through a beam splitter $BS_{1}$ and a
mirror $M$ as shown in Figure 13. The beam reflected from the mirror
has its polarization
rotated by $90$ degrees before passage through a second beam splitter $%
BS_{2} $, where quantum interference takes place.

\

\

\begin{figure}[!t]
\centerline{\epsfxsize=3.5in \epsfbox{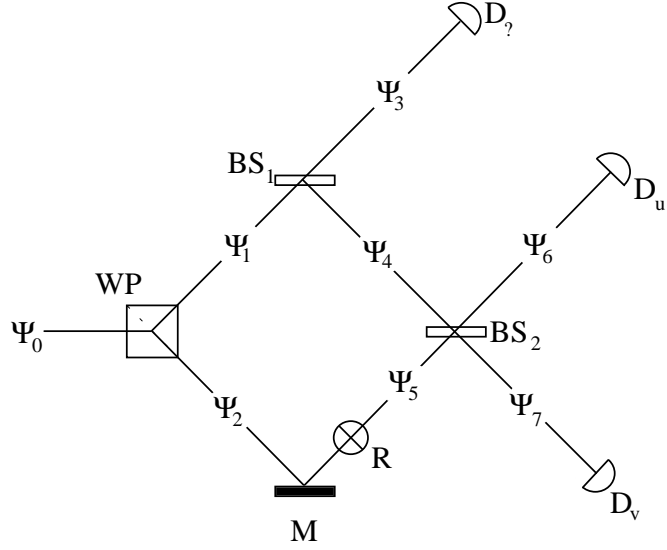}}
\caption{An interference experiment requiring a POVM description.}
\end{figure}

In this experiment, the initial state is given as a superposition of two
non-orthogonal states,\qquad\
\begin{equation}
|\Psi _{0}\rangle =\alpha |u\rangle +\beta |v\rangle
\end{equation}
where $\langle u|v\rangle =\cos \theta$. The calculations in \cite
{BRANDT-99,BRANDT-02} give the following set of POVM operators:
\begin{equation}
E_{u}=\frac{\Bbb{I}_{\mathcal{H}}-|v\rangle \langle v|}{1+\cos \theta }%
,\;\;\;E_{v}=\frac{\Bbb{I}_{\mathcal{H}}-|u\rangle \langle u||}{1+\cos
\theta },\;\;\;E_{?}=\Bbb{I}_{\mathcal{H}}-E_{u}-E_{v}.  \label{FFF}
\end{equation}
The outcome probabilities are found to be
\begin{eqnarray}
P\left( u|\Psi _{0}\right) &=&\langle \Psi _{0}|E_{u}|\Psi _{0}\rangle
=|\alpha |^{2}(1-\cos \theta )  \nonumber \\
P\left( v|\Psi _{0}\right) &=&\langle \Psi _{0}|E_{v}|\Psi _{0}\rangle
=|\beta |^{2}(1-\cos \theta ) \\
P\left( ?|\Psi _{0}\right) &=&\langle \Psi _{0}|E_{?}|\Psi _{0}\rangle
=|\alpha +\beta |^{2}\cos \theta .  \nonumber
\end{eqnarray}

The quantum register description involves eight qubits, if we ignore any
phase shift at the mirror. \

\

\begin{figure}[!t]
\centerline{\epsfxsize=3.5in \epsfbox{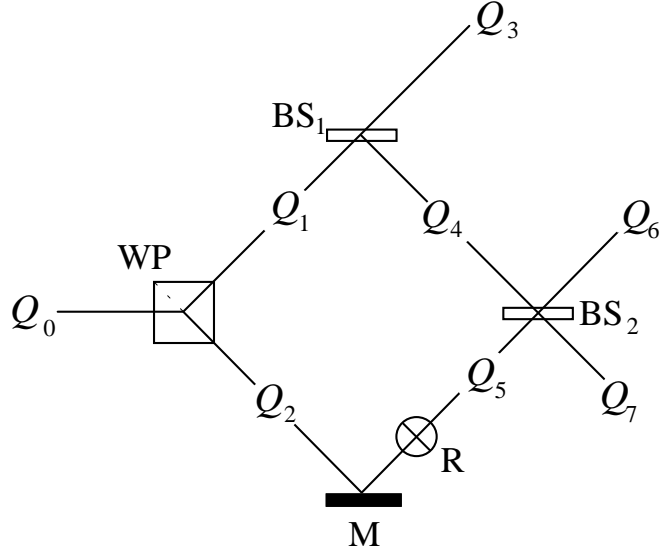}}
\caption{Qubit assignment for the experiment shown in Figure 13.}
\end{figure}

The quantum register calculation goes as follows:

\subparagraph{i) Wollaston prism $WP$:}

\begin{equation}
\Bbb{A}_{0}^{+}\rightarrow \left( \alpha +\beta \right) \cos (\frac{_{1}}{%
^{2}}\theta )\Bbb{A}_{1}^{+}+\left( \alpha -\beta \right) \sin (\frac{_{1}}{%
^{2}}\theta )\Bbb{A}_{2}^{+},
\end{equation}

\subparagraph{ii) Beam splitter $BS_{1}$:}

\begin{equation}
\Bbb{A}_{1}^{+}\rightarrow \sqrt{1-\tan ^{2}(\frac{_{1}}{^{2}}\theta )}\Bbb{A%
}_{3}^{+}+i\tan (\frac{_{1}}{^{2}}\theta )\Bbb{A}_{4}^{+},
\end{equation}

\subparagraph{iii) Mirror $M$ and $90^{\circ }$ Polarization Rotation $R$:}

\begin{equation}
\Bbb{A}_{2}^{+}\rightarrow -\Bbb{A}_{5}^{+},
\end{equation}

\subparagraph{iv) Beam splitter $BS_{2}$:}

\begin{eqnarray}
\Bbb{A}_{4}^{+} &\rightarrow &\frac{i}{\sqrt{2}}\Bbb{A}_{6}^{+}+\frac{1}{%
\sqrt{2}}\Bbb{A}_{7}^{+},  \nonumber \\
\Bbb{A}_{5}^{+} &\rightarrow &\frac{1}{\sqrt{2}}\Bbb{A}_{6}^{+}+\frac{i}{%
\sqrt{2}}\Bbb{A}_{7}^{+}.
\end{eqnarray}
Hence finally,
\begin{equation}
\Bbb{A}_{0}^{+}\rightarrow \left( \alpha +\beta \right) \sqrt{\cos \theta }%
\Bbb{A}_{3}^{+}-\alpha \sqrt{1-\cos \theta }\Bbb{A}_{6}^{+}+i\beta \sqrt{%
1-\cos \theta }\Bbb{A}_{7}^{+},  \label{EEE}
\end{equation}
i.e.
\begin{eqnarray}
|\Psi _{in})\equiv |1)\rightarrow |\Psi _{out}) &=&\left( \alpha +\beta
\right) \sqrt{\cos \theta }|2^{3})  \nonumber \\
&&-\alpha \sqrt{1-\cos \theta }|2^{6})+i\beta \sqrt{1-\cos \theta }|2^{7}).
\end{eqnarray}
This gives the conditional probabilities
\begin{eqnarray}
P\left( ?|\Psi _{0}\right) &\equiv &|(2^{3}|\Psi _{out})|^{2}=|\alpha +\beta
|^{2}\cos \theta  \nonumber \\
P\left( u|\Psi _{0}\right) &=&|2^{6}|\Psi _{out})|^{2}=|\alpha |^{2}\left(
1-\cos \theta \right) \\
P\left( v|\Psi _{0}\right) &=&|2^{7}|\Psi _{out})|^{2}=|\beta |^{2}\left(
1-\cos \theta \right)  \nonumber
\end{eqnarray}
exactly as in the conventional description.

It is here that the advantage of working with the quantum register
description begins to show itself. The transition rule (\ref{EEE}) not only
has all the hallmarks of the PVM description (albeit in a Hilbert space of
dimension $2^{8})$, but is conceptually more understandable than the set of
POVM operators (\ref{FFF}). In particular, all of the detector qubits $%
\mathcal{Q}_{3}$, $\mathcal{Q}_{6}$ and $\mathcal{Q}_{7}$ are treated
in the same way, whereas the status of the detector labelled ``?'' is
considered different to the detectors labelled ``u" and ``v". The
quantum register description of each stage of the experiment makes it
clear that the original formulation of the experiment in terms of
non-orthogonal basis vectors is somewhat contrived and strictly
speaking, not necessary.

\section{Interpretation of higher rank states}

There are certain situations in quantum register physics where states of
rank higher than one are encountered. We discuss some of these next.

\subsection{Independent experiments}

Suppose two Stern-Gerlach experiments are performed separately, completely
independently of each other in different parts of the world. In such a case,
we can describe the two experiments by a single rank-$6$ quantum register
with rank-2 states, as shown in Figure 15.

\

\

\begin{figure}[!t]
\centerline{\epsfxsize=2.5in \epsfbox{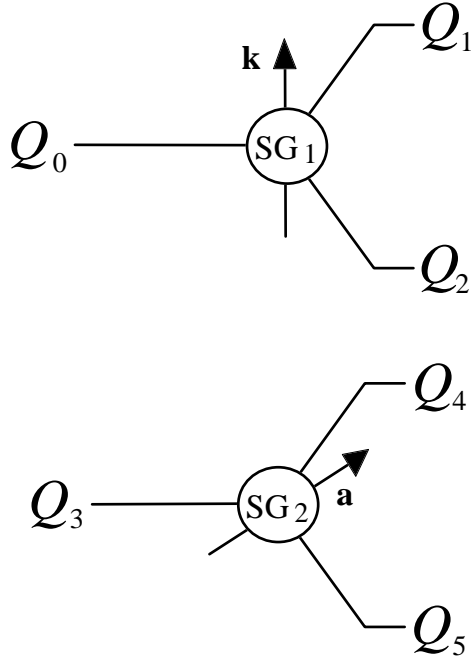}}
\caption{Qubit assignment for two independent Stern-Gerlach experiments.}
\end{figure}

In this case, the initial state is given by
\begin{equation}
|\Psi _{in})=\Bbb{A}_{0}^{+}\Bbb{A}_{3}^{+}|0)=|100100)=|2^{0}+2^{3})=|9).
\end{equation}
If each experiment is truly independent, then we can write
\begin{eqnarray}
\Bbb{A}_{0}^{+} &\rightarrow &\alpha \Bbb{A}_{1}^{+}+\beta \Bbb{A}%
_{2}^{+},\;\;\;\;\;|\alpha |^{2}+|\beta |^{2}=1,  \nonumber \\
\Bbb{A}_{3}^{+} &\rightarrow &\gamma \Bbb{A}_{4}^{+}+\delta \Bbb{A}%
_{5}^{+},\;\;\;\;\;|\gamma |^{2}+|\delta |^{2}=1,
\end{eqnarray}
so
\begin{eqnarray*}
|\Psi _{in}) &\rightarrow &|\Psi _{out}\rangle =\left( \alpha \Bbb{A}%
_{1}^{+}+\beta \Bbb{A}_{2}^{+}\right) \left( \gamma \Bbb{A}_{4}^{+}+\delta
\Bbb{A}_{5}^{+}\right) |0) \\
&=&|\psi )_{1}\otimes |\phi )_{2},
\end{eqnarray*}
where
\begin{eqnarray}
|\psi )_{1} &\equiv &\alpha |0)_{0}|1)_{1}|0)_{2}+\beta
|0)_{0}|0)_{1}|1)_{2},  \nonumber \\
|\phi )_{2} &\equiv &\gamma |0)_{3}|1)_{4}|0)_{5}+\delta
|0)_{3}|0)_{4}|1)_{5}.
\end{eqnarray}
In other words, independent experiments are modelled in quantum register
physics by separable states of rank higher than unity.

\subsection{Change of rank experiments: (EPR)}

Experiments of the type discussed by Einstein, Podolsky and Rosen \cite{EPR}
cause conceptual problems because they invoke quantum non-locality. However,
we note that non-locality is already inherent in real experiments (recall
the two distinct spots in the Stern-Gerlach experiment, shown in Figure 1).
Therefore, what is conventionally regarded as non-locality is really a
matter of scale or degree.

Suppose we prepared a spin-zero bound state of an electron and a positron,
given in the conventional description by
\begin{equation}
|\Psi \rangle =\frac{1}{\sqrt{2}}\left\{ |+\mathbf{k\rangle }_{-}\otimes |-%
\mathbf{k\rangle }_{+}-|-\mathbf{k\rangle }_{-}\otimes |+\mathbf{k\rangle }%
_{+}\right\} .
\end{equation}
Alice and Bob are two well-separated observers, each with their own particle
species filters and Stern-Gerlach equipment. Alice can detect and test for
electron spin only, whereas Bob can detect and test for positron spin only.
Alice sets her quantization axis along $\mathbf{k}=(0,0,1)$, whereas Bob
sets his along direction
\begin{equation}
\mathbf{a}=(\sin \theta \cos \phi ,\sin \theta \sin \phi ,\cos \theta )
\end{equation}
Now whenever Alice finds an electron passes through her apparatus with spin $%
|+\mathbf{k\rangle }$, Bob will find his positron passes through either of
the $|+\mathbf{a\rangle }$ or $|-\mathbf{a\rangle }$ channels in a random
way, with frequency given correctly by quantum mechanics.

The quantum register description requires five qubits, as shown in Figure 16:

\

\

\begin{figure}[!t]
\centerline{\epsfxsize=4.0in \epsfbox{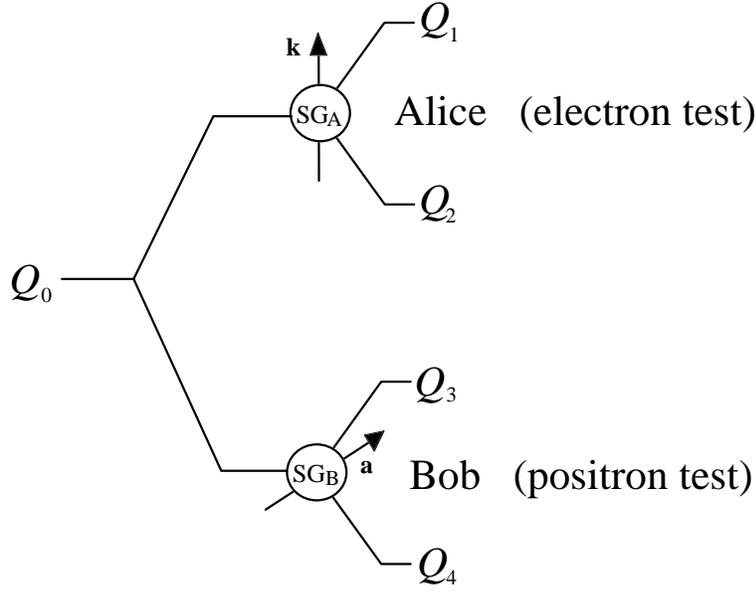}}
\caption{Qubit assignment for an EPR experiment.}
\end{figure}

The qubit assignment is

$\mathcal{Q}_{0}$: initial spinless bound state,

$\mathcal{Q}_{1}$: electron spin $|+\mathbf{k}\rangle ,$ $\;\;\mathcal{Q}%
_{2} $: electron spin $|-\mathbf{k}\rangle $

$\mathcal{Q}_{3}$: positron spin $|+\mathbf{a}\rangle ,\;\;\;\mathcal{Q}_{4}$%
: positron spin $|-\mathbf{a}\rangle $.

A conventional quantum mechanics calculation permits us to write
\[
|\Psi \rangle =\frac{1}{\sqrt{2}}\left\{
\begin{array}{c}
\sin (\frac{_{1}}{^{2}}\theta )e^{-i\phi }|+\mathbf{k\rangle }_{-}|+\mathbf{%
a\rangle }_{+}+\cos (\frac{_{1}}{^{2}}\theta )e^{-i\phi }|+\mathbf{k\rangle }%
_{-}|-\mathbf{a\rangle }_{+} \\
-\cos (\frac{_{1}}{^{2}}\theta )|-\mathbf{k\rangle }_{-}|+\mathbf{a\rangle }%
_{+}+\sin (\frac{_{1}}{^{2}}\theta )|-\mathbf{k\rangle }_{-}|-\mathbf{%
a\rangle }_{+}
\end{array}
\right\},
\]
so we deduce
\begin{eqnarray*}
\Bbb{A}_{0}^{+} &\rightarrow &\frac{\sin (\frac{_{1}}{^{2}}\theta )e^{-i\phi
}}{\sqrt{2}}\Bbb{A}_{1}^{+}\Bbb{A}_{3}^{+}+\frac{\cos (\frac{_{1}}{^{2}}%
\theta )e^{-i\phi }}{\sqrt{2}}\Bbb{A}_{1}^{+}\Bbb{A}_{4}^{+} \\
&&-\frac{\cos (\frac{_{1}}{^{2}}\theta )}{\sqrt{2}}\Bbb{A}_{2}^{+}\Bbb{A}%
_{3}^{+}+\frac{\sin (\frac{_{1}}{^{2}}\theta )}{\sqrt{2}}\Bbb{A}_{2}^{+}\Bbb{%
A}_{4}^{+}.
\end{eqnarray*}
Hence the initial state
\[
|\Psi _{in})=|1)=\Bbb{A}_{0}^{+}|0)
\]
is a rank-one state which changes into an entangled rank-two state.

\subsection{Two-particle interferometry}

In 1989, Horne, Shimony and Zeilinger discussed an experiment where an
entangled two-photon state passes through the device shown in the quantum
register representation, Figure 17 \cite{HSZ-89}. $M_{1}$, $M_{2}$, $M_{3}$
and $M_{4}$ are mirrors, $\phi _{1}$ and $\phi _{2}$ are variable phase
shifts, and $BS_{1}$ and $BS_{2}$ are beam splitters. Qubits $\mathcal{Q}%
_{7},\mathcal{Q}_{8}$, $\mathcal{Q}_{9}$ and $\mathcal{Q}_{10}$ are
associated with photon detectors. The quantities of interest are the
two-particle coincidence count rates and their dependence on the phase-shift
angles $\phi _{1}$, $\phi _{2}$, which can be varied at will throughout the
experiment.

\

\

\begin{figure}[!t]
\centerline{\epsfxsize=4.0in \epsfbox{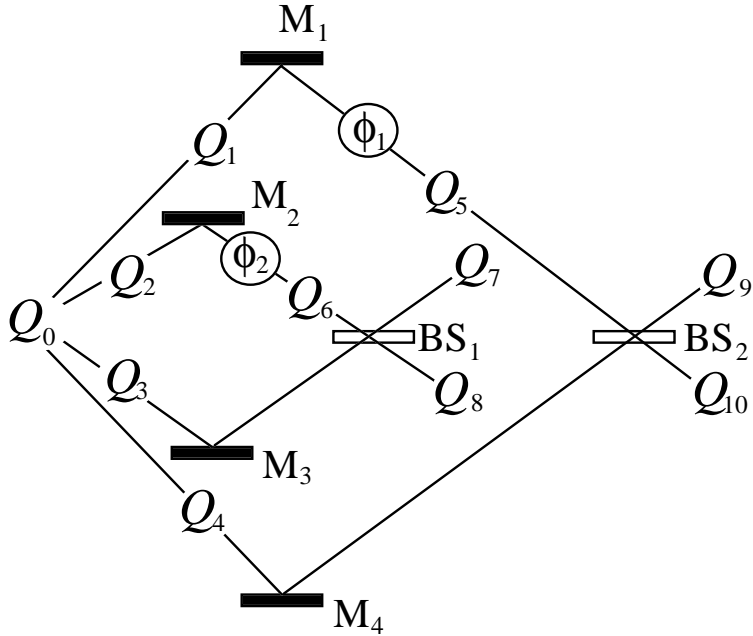}}
\caption{Qubit assignment for a two-particle interferometry experiment.}
\end{figure}

The conventional representation of the initial state is
\begin{equation}
|\Psi _{in}\rangle =\frac{1}{\sqrt{2}}\left\{ |\mathbf{k}_{A}\rangle _{1}|%
\mathbf{k}_{C}\rangle _{2}+|\mathbf{k}_{D}\rangle _{1}|\mathbf{k}_{B}\rangle
_{2}\right\} ,
\end{equation}
where the wave vectors $\mathbf{k}_{A}$, $\mathbf{k}_{B}$, $\mathbf{k}_{C}$
and $\mathbf{k}_{D}$ are identified with qubits $\mathcal{Q}_{1}$, $\mathcal{%
Q}_{2}$, $\mathcal{Q}_{3}$ and $\mathcal{Q}_{4}$ respectively, and
subscripts $1$ and $2$ refer to the two particles involved.

Details of the conventional calculation are not given here. The quantum
register account goes as follows. The initial state is of rank one,
regardless of the fact that it consists of an entangled two-photon state.
There is only one source, which means the prepared state is of rank
one. However, because two photons can be detected simultaneously at
independent sites subsequent to state preparation, the initial state
changes rank to a rank-two state. This is represented by the operator
transition

\begin{equation}
\Bbb{A}_{0}^{+}\rightarrow \frac{1}{\sqrt{2}}\left\{ \Bbb{A}_{1}^{+}\Bbb{A}%
_{3}^{+}+e^{i\theta }\Bbb{A}_{2}^{+}\Bbb{A}_{4}^{+}\right\} ,
\end{equation}
where the angle $\theta $ depends on the detailed placement of the various
pieces of equipment, as discussed in \cite{HSZ-89}. We ignore any overall
changes of phase due to the mirrors, as this will not affect probabilities.
The effect of the phase shifts $\phi _{1},\phi _{2}$ gives
\begin{equation}
\Bbb{A}_{1}^{+}\rightarrow e^{i\phi _{1}}\Bbb{A}_{5}^{+},\;\;\;\Bbb{A}%
_{2}^{+}\rightarrow e^{i\phi _{2}}\Bbb{A}_{6}^{+}
\end{equation}
and finally, the beam splitters give the transitions
\begin{eqnarray}
\Bbb{A}_{6}^{+} &\rightarrow &\frac{1}{\sqrt{2}}\left\{ \Bbb{A}_{8}^{+}+i%
\Bbb{A}_{7}^{+}\right\} ,\;\;\;\Bbb{A}_{3}^{+}\rightarrow \frac{1}{\sqrt{2}}%
\left\{ \Bbb{A}_{7}^{+}+i\Bbb{A}_{8}^{+}\right\}  \nonumber \\
\Bbb{A}_{4}^{+} &\rightarrow &\frac{1}{\sqrt{2}}\left\{ \Bbb{A}_{9}^{+}+i%
\Bbb{A}_{10}^{+}\right\} ,\;\;\;\Bbb{A}_{5}^{+}\rightarrow \frac{1}{\sqrt{2}}%
\left\{ \Bbb{A}_{10}^{+}+i\Bbb{A}_{9}^{+}\right\} .
\end{eqnarray}
This is all that is required for the complete quantum register calculation.
We find
\begin{eqnarray}
\Bbb{A}_{0}^{+} &\rightarrow &\frac{1}{2\sqrt{2}}\left\{ e^{i\phi
_{1}}-e^{i(\theta +\phi _{2})}\right\} \Bbb{A}_{7}^{+}\Bbb{A}_{10}^{+}+\frac{%
1}{2\sqrt{2}}\left\{ ie^{i\phi _{1}}+ie^{i(\theta +\phi _{2})}\right\} \Bbb{A%
}_{7}^{+}\Bbb{A}_{9}^{+}  \nonumber \\
&&+\frac{1}{2\sqrt{2}}\left\{ ie^{i\phi _{1}}+ie^{i(\theta +\phi
_{2})}\right\} \Bbb{A}_{8}^{+}\Bbb{A}_{10}^{+}+\frac{1}{2\sqrt{2}}\left\{
-e^{i\phi _{1}}+e^{i(\theta +\phi _{2})}\right\} \Bbb{A}_{8}^{+}\Bbb{A}%
_{9}^{+}
\end{eqnarray}
for the full experiment. The two-particle co-incidence probabilities are
found to be
\begin{eqnarray}
P\left( 7\&9|\Psi _{in}\right) &=&\frac{1}{4}\left\{ 1+\cos \left( \theta
+\phi _{2}-\phi _{1}\right) \right\} ,  \nonumber \\
P\left( 7\&10|\Psi _{in}\right) &=&\frac{1}{4}\left\{ 1-\cos \left( \theta
+\phi _{2}-\phi _{1}\right) \right\} ,  \nonumber \\
P\left( 8\&9|\Psi _{in}\right) &=&\frac{1}{4}\left\{ 1-\cos \left( \theta
+\phi _{2}-\phi _{1}\right) \right\} , \\
P\left( 8\&10|\Psi _{in}\right) &=&\frac{1}{4}\left\{ 1+\cos \left( \theta
+\phi _{2}-\phi _{1}\right) \right\} ,  \nonumber
\end{eqnarray}
in precise agreement with the calculation of Horne, Shimony and Zeilinger
\cite{HSZ-89}, assuming no losses in the system.

\subsection{Other scenarios involving higher rank states}

Obvious candidate experiments for future discussion are $i)$ interference of
photons from different sources, $ii)$ teleportation and $iii)$ experiments
where a sequence of wave-pulses is set up moving towards target detectors.
In such cases, it is possible that the source apparatus gets destroyed
before the detectors register anything. This happens in
astrophysics, where it is quite normal for astronomers to receive light from
sources which have long ceased to exist.

Many interesting physical ideas remain to be explored. Of greatest interest
to us is the possibility of modelling physical space as a quantum register
of enormous, possible infinite rank. Work is in hand currently on this front.

Perhaps the ultimate development of quantum register physics would be to
provide an account of quantum cosmology, which we could call quantum
register cosmology \cite{E+J:04}. Such a theory would be the
ultimate vindication of Feynman's vision of physics
simulated in terms of computation. We can only speculate at this time
as to the details of such a theory . In any programme attempting to
extending the quantum register description to the universe, we would
have to face great conceptual issues as well as technical problems.
It is not accepted universally that quantum mechanics can be applied
to the universe considered as a system, for instance. Certainly, such
a concept would require an endophysical account rather than the
exophysical one we have been forced to use thus far. But this is
precisely what Feynman was saying in the quote we gave at the start
of this paper.

We believe that Feynman was not just advocating an endophysical vision of
physics. What he was referring to would, in our view, inevitably lead to
quantum register cosmology. In other words, a quantum computational theory
of everything.

\

\textbf{Acknowledgements}

\

I am indebted to many people for the numerous discussions which played an
essential role in the formulation of the ideas behind this paper. Of these,
I have to mention Lino Buccheri, Mark Stuckey, Jason Ridgway-Taylor, and
most particularly, Jon Eakins.


\newpage

\begin{thebibliography}{99}
\bibitem{FEYNMAN-82}  Richard~P. Feynman. \newblock Simulating physics with
computers. \newblock {\em Int. Journal. Theor. Phys.}, 21(6/7):
467--488, 1982.

\bibitem{WU+LIDAR-02}  L.~A. Wu and D.~A. Lidar. \newblock Qubits as
parafermions. \newblock {\em Journal of Mathematical Physics},
43(9): 4506--4525, 2002.

\bibitem{JORDAN+WIGNER-28}  P.~Jordan and E.~P. Wigner. \newblock {\em Z.
Physik}, 47: 631, 1928.

\bibitem{STERN-GERLACH-22A}  W.~Gerlach and O.~Stern. \newblock {\em Z.
Phys.}, 8: 110, 1922.

\bibitem{STERN-GERLACH-22B}  W.~Gerlach and O.~Stern. \newblock {\em Z.
Phys.}, 9: 349, 1922.

\bibitem{VON-NEUMANN:55}  J.~Von Neumann. \newblock {\em The Mathematical
Foundations of Quantum Mechanics}. \newblock Princeton University Press,
1955.

\bibitem{PERES:93}  Asher Peres. \newblock {\em Quantum Theory: Concepts and
Methods}. \newblock Kluwer Academic Publishers, 1993.

\bibitem{ZEILINGER-81}  A.~Zeilinger. \newblock General properties of
lossless beam splitters in interferometry. \newblock {\em Am. J. Phys.},
49(9): 882--883, 1981.

\bibitem{BRANDT-99}  Howard~E. Brandt. \newblock Positive operator valued
measure in quantum information processing. \newblock {\em Am. J. Phys.},
67(5): 434--439, 1999.

\bibitem{BRANDT-02}  Howard~E. Brandt. \newblock Quantum measurement with a
positive operator-valued measure. \newblock {\em Wigner Centennial
Conference, Pecs, Hungary, 8-12 July}, 2002.

\bibitem{EPR}  B.~Podolsky A.~Einstein and N.~Rosen. \newblock Can
quantum-mechanical description of physical reality be considered  complete? %
\newblock {\em Phys. Rev.}, 47: 777--780, 1935.

\bibitem{HSZ-89}  Abner~Shimony Michael A.~Horne and Anton Zeilinger. %
\newblock Two-particle interferometry. \newblock {\em Phys. Rev. Lett.},
62(19): 2209--2212, 1989.

\bibitem{E+J:04}  J. Eakins and G. Jaroszkiewicz, \emph{Quantum Register
Cosmology}, in ``Progress in General relativity and Quantum Cosmology", Nova
Science Publishers, Inc (in press, 2004).
\end{thebibliography}
\end{document}